\documentclass[RNAAS]{aastex62}

\begin{document}

\title{Mid-infrared variability of the neutrino source blazar TXS\,0506$+$056}

%% Note that the corresponding author command and emails has to come
%% before everything else. Also place all the emails in the \email
%% command instead of using multiple \email calls.
\correspondingauthor{Krisztina \'Eva Gab\'anyi}
\email{krisztina.g@gmail.com}

\author{Krisztina \'Eva Gab\'anyi}
\altaffiliation{}
\affiliation{MTA-ELTE Extragalactic Astrophysics Research Group, P\'azm\'any P\'eter s\'et\'any 1/A, H-1117 Budapest, Hungary}
\affiliation{Konkoly Observatory, MTA CSFK, Konkoly-Thege Mikl\'os \'ut 15-17, H-1121 Budapest, Hungary}

\author{Attila Mo\'or}
\affiliation{Konkoly Observatory, MTA CSFK, Konkoly-Thege Mikl\'os \'ut 15-17, H-1121 Budapest, Hungary}

\author{S\'andor Frey}
\affiliation{Konkoly Observatory, MTA CSFK, Konkoly-Thege Mikl\'os \'ut 15-17, H-1121 Budapest, Hungary}

%% Note that RNAAS manuscripts DO NOT have abstracts.

%% Start the main body of the article. If no sections in the 
%% research note leave the \section call blank to make the title.
\section{} 

The IceCube instrument detected a high-energy cosmic neutrino event (IceCube\_170922A) on 2017 September 22 \citep{sci1}. The electromagnetic follow-up campaigns associated the event with the flaring $\gamma$-ray blazar TXS\,0506$+$056 \citep{Padovani2018}. Investigation of the archival data of the IceCube instrument revealed an excess of high-energy neutrinos in the direction of TXS\,0506$+$056 compared to the atmospheric background between 2014 September and 2015 March \citep{sci2} possibly also associated with the blazar \citep{Padovani2018}. Blazars are radio-loud active galactic nuclei where one of the synchrotron-emitting jets is pointed at small angle to line of sight, thus leading to relativistically boosted emission. They are thought to be promising sources of cosmic neutrinos \citep[e.g., ][]{mannheim}. %\citep[for a review  see][and references therein]{Padovani2018}

To investigate the mid-infrared (MIR) variability of TXS\,0506$+$056, we analyzed the data collected by the {\it Wide-Field Infrared Survey Explorer} \citep[{\it WISE}, ][]{wise} satellite. {\it WISE} scanned the whole sky at four bands, at $3.4$, $4.6$, $12$, and $22\mu$m (W1, W2, W3, and W4, respectively) during 2010. After the end of the original mission, the survey continued as the NEOWISE \citep[Near-Earth Object WISE, ][]{neowise} project, and after a 3-year hibernation gap as the NEOWISE Reactivation Mission. Since the cooling material required for the detectors W3 and W4 were depleted, latter measurements were conducted only at the W1 and W2 bands. We downloaded {\it WISE} single exposure data\footnote{\url{http://irsa.ipac.caltech.edu/Missions/wise.html}} of TXS\,0506$+$056. To avoid potentially unreliable data, we followed the outline in the Explanatory Supplement\footnote{\url{http://wise2.ipac.caltech.edu/docs/release/allwise/expsup/sec3\_2.html}} and discarded photometry associated with bad quality frames (`qual\_frame'$=0$). The observations of TXS\,0506$+$056 were not performed too close to the boundary of the South Atlantic Anomaly, i.e., this issue does not affect us. With respect to the possible contamination by scattered moonlight, we examined the photometry of $63$ {\it WISE} sources within $30\arcmin$ of TXS\,0506$+056$ with similar brightnesses ($\pm 1^\mathrm{m}$) in W1 and W2 bands. We found that data points flagged due to this effect are well consistent with the neighbouring non-flagged data, thus we retained them for further analysis. The obtained light curves of TXS\,0506$+$056 in W1 and W2 bands consist of 10 mission phases taken between 2010 March and 2017 September. Each mission phase covers $\approx 1-5$ days with a dozen measurement points.

\begin{figure}[h!]
\begin{center}
\includegraphics[width=\columnwidth,angle=0]{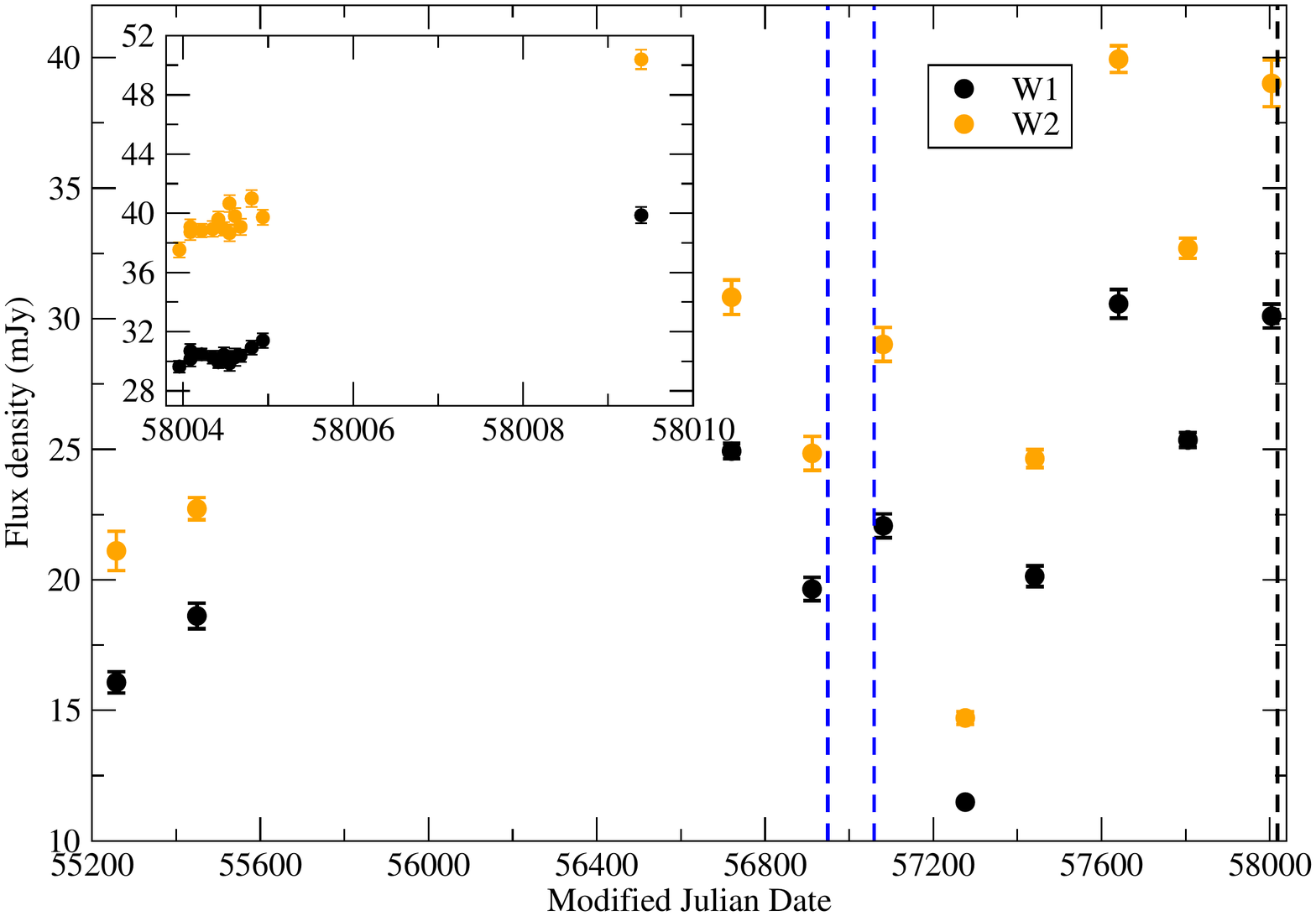}
\caption{WISE light curve of TXS\,0506$+$056, black and orange symbols are for band W1 and W2, respectively. Points are averages of each mission phases. The error bars represent the variability within the given phase. Blue vertical lines mark the excess neutrino phase reported in \cite{sci2}. Black vertical line marks the IceCube\_170922A event. The inset shows a zoom-in to the last mission phase. Here the error bars represent the formal errors of the single exposures. \label{fig:lc}}
\end{center}
\end{figure}

TXS\,0506$+056$ shows significant MIR variability during year-long time scale (Fig. \ref{fig:lc}) with more than a factor of $2.5$ peak-to-peak variation in both bands. A brightening can be seen in the last two mission phases, which is concurrent with the $\gamma$-ray flare started in April 2017 and peaked around the IceCube\_170922A event. No {\it WISE} observation is available during the previous excess neutrino event, however TXS\,0506$+$056 became brighter in MIR after that period. 

We used two indices, the correlation-based Stetson index \citep{stetson} and the $\chi^2$ test to assess the variability behavior on day-long time scales 
in each individual mission phase. The threshold values of these were determined based on {\it WISE} data of neighbouring sources with similar brightnesses \citep{pos}. Two mission phases are found to show significant short time scale changes. In the first mission phase (at MJD $55258$), a fading can be seen, reaching $\sim 10$\,\% in W2. In the last mission phase preceding the IceCube\_170922A event by a few days, a MIR brightening (Fig. \ref{fig:lc} inset) as high as $\sim 30$\,\% in both bands occurred during $4$ days. Similar variability behaviour and intraday variability in MIR bands were found in $\gamma$-ray bright radio-loud narrow-line Seyfert 1 sources, which are thought to be blazar-like active galactic nuclei with a relatively small black hole mass \citep[$10^6-10^8 M_\mathrm{BH}$,][]{yuan2008}. The short time-scale implies small emitting region indicating that instead of star formation in the host galaxy, or the torus of the active galactic nucleus, the jet is likely to be responsible for the variability \citep{2012Jiang, 2018Yang, pos}.

%In the last mission phase preceding the IceCube\_170922A event by a few days, a substantial MIR brightening occurred (Fig. \ref{fig:lc} inset) possibly related to the $\gamma$-ray flare. 
%In the first mission phase (at MJD $55258$), a fading can be seen, reaching $\sim 10$\,\% in W2. Similar intraday MIR variability was found in $\gamma$-ray bright radio-loud narrow-line Seyfert 1 sources, which are thought to be blazar-like active galactic nuclei with a relatively small black hole mass \citep[$10^6-10^8 M_\mathrm{BH}$,][]{yuan2008}.

%% An example figure call using \includegraphics

\acknowledgments

K\'EG acknowledges the J\'anos Bolyai Research Scholarship of the Hungarian Academy of Sciences. This work was supported by the NKFIH-OTKA NN110333 grant. This publication makes use of data products from the WISE, which is a joint project of the University of California, Los Angeles, and the Jet Propulsion Laboratory/California Institute of Technology, funded by the National Aeronautics and Space Administration.

\end{document}